\documentclass[reprint,superscriptaddress,showpacs,amsmath,amssymb,aps,prl]{revtex4-1}
\usepackage{gensymb}
\usepackage{graphicx}
\usepackage{dcolumn}
\usepackage{bm}
\usepackage{hyperref}

\begin{document}


\title{Transition radiation in photonic topological crystals: quasi-resonant excitation of robust edge states by a moving charge}

\author{Yang Yu}
\affiliation{School of Applied and Engineering Physics, Cornell University, Ithaca, NY 14853, USA}
\author{Kueifu Lai}
\affiliation{Department of Physics, University of Texas at Austin, Austin, TX 78712, USA.}
\affiliation{School of Applied and Engineering Physics, Cornell University, Ithaca, NY 14853, USA}
\author{Jiahang Shao}\author{John Power}\author{Manoel Conde}\author{Wanming Liu}\author{Scott Doran}\author{Chunguang Jing}\author{Eric Wisniewski}
\affiliation{Argonne National Laboratory, Lemont, Illinois 60439, USA}
\author{Gennady Shvets}
 \email{gshvets@cornell.edu}
\affiliation{School of Applied and Engineering Physics, Cornell University, Ithaca, NY 14853, USA}

\date{\today}

\begin{abstract}
We demonstrate, theoretically and experimentally, that a traveling electric charge passing from one photonic crystal into another generates edge waves -- electromagnetic modes with frequencies inside the common photonic bandgap localized at the interface -- via a process of transition edge-wave radiation (TER). A simple and intuitive expression for the TER spectral density is derived and then applied to a specific structure: two interfacing photonic topological insulators with opposite spin-Chern indices. We show that TER breaks the time-reversal symmetry and enables valley- and spin-polarized generation of topologically protected edge waves propagating in one or both directions along the interface. Experimental measurements at the Argonne Wakefield Accelerator Facility are consistent with the excitation and localization of the edge waves. The concept of TER paves the way for novel particle accelerators and detectors.
\end{abstract}

\maketitle

Generation of electromagnetic (EM) waves by moving electric charges is one of the most fundamental phenomena in physics. While a charge must be accelerated to produce EM radiation in free space, this requirement no longer exists in optically-dense media.  Even in a homogeneous isotropic medium, the Cherenkov radiation (CR) \cite{CR} by a charge travelling with a constant velocity $v$ can be produced when the phase velocity $v_{\rm ph}$ of EM waves is smaller than $v$. In an inhomogeneous medium, transition radiation (TR) \cite{first} -- usually studied in the context of a charge crossing an interface between two media with different permittivities and/or permeabilities -- can also be produced by a constant-velocity motion \cite{mik,frank70,gar,ginz,PhysRevLett.67.2962,PhysRevD.10.3594,wartski1975interference,adamo2012electron} regardless of the magnitude of $v$. TR has already found numerous applications in particle detectors and beam diagnostics \cite{wartski1975interference,PhysRevD.12.1289}. More recently, there has been considerable interest in expanding the TR concept to more complex geometries and structures, including the resonant transition radiation~\cite{PhysRevA.40.1918,finman_83,bekefi_86,lin2018controlling} in multi-interfacial materials that form a one-dimensional (1D) photonic crystal. TR has also been used to excite surface plasmon polaritons (SPPs) \cite{ginz,kuttge2009local} and guided modes in thin films \cite{chen1975detection,yurtsever2008formation}, which are hard to be directly excited by far-field (e.g., laser) radiation.

\begin{figure}
\includegraphics[width=\linewidth]{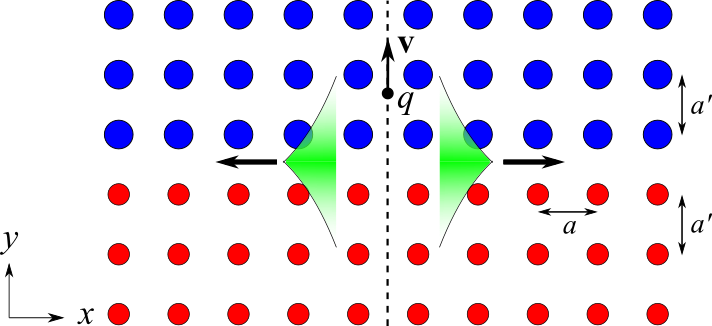}
\caption{\label{fig:schematic} A schematic of transition radiation by a point charge at the interface of two photonic crystals. The charge moves from one PhC (small/red circles) to another (large/blue circles) with constant velocity $v$. Guided (edge) modes (green shades) propagating in the $x$ direction are excited with frequencies inside the shared bandgap of the two PhCs , as well as bulk modes (not shown) at frequencies outside the bandgap. The period $a$ along $x$ direction and the lattice period $a'$ along the beam's path are labelled.}
\end{figure}

The key limitation of all these approaches to producing TR is that fast charged particles must be sent through a solid medium, resulting in rapid energy loss by the electrons, as well as the inevitable incoherent emission \cite{de2010optical}. For example, a $1$ MeV electron loses all of its energy after propagating through just under $3$ mm of silicon. Charging of multi-layer dielectric structures bombarded by high-charge bunches also limits their longevity \cite{wilson2013electron}. Therefore, one is led to consider an intriguing yet unexplored possibility of producing TR in a photonic crystal (PhC) designed to have an empty region that provides an unobstructed path for the moving charge (see Fig.~\ref{fig:schematic}). However, the physics of TR excitation in two- and three-dimensional periodic media has not been studied either theoretically or experimentally, with a few exception of 1D multilayer films~\cite{PhysRevA.40.1918,bekefi_86,lin2018controlling}. Even in those studies, the emphasis was on the excitation of the modified Cherenkov (i.e. bulk) radiation, and the feasibility of sending electrons through solid medium was assumed. In this Letter, we extend the concept of TR to the case of a charge crossing the interface between two PhCs and emitting {\it guided waves} that are localized to the interface. In particular, we consider the previously unexplored concept of TR into topologically protected edge waves (TPEWs) that exist at the domain wall between two topologically-distinct photonic topological insulators (PTIs) \cite{hafezi2011robust,khanikaev2013photonic,hafezi2013imaging,SciRep}. We demonstrate that the moving charge breaks the time-reversal symmetry of TPEWs and enables spin- and valley-polarized emission of TPEWs that are routed into spin-locked ports. In condensed matter physics it has been shown that circularly polarized light can excite spin-locked currents on the surface of topological insulators \cite{soifer2017band}. Among practical attractions of TPEWs are their one-dimensional (i.e. localized in the other two dimensions) nature, and the ability for reflection-free propagation around sharp corners \cite{PhysRevLett.114.127401}. Similarly to SPPs, TPEWs cannot directly couple to bulk EM waves. However, SPPs can be also excited by moving charges via the CR mechanism \cite{de2010optical} because their dispersion curves are below the light line ($v_{\rm ph} < c$) due to their polaritonic nature, while TPEWs frequently cannot be because the phase velocities of the guided EM waves typically satisfy $v_{\rm ph} > c$.

We start by developing a general formalism of guided waves' excitation by a TR mechanism as illustrated in Fig.~\ref{fig:schematic}, where a point electrical charge $q$ is shown moving uniformly with velocity $v$, crossing the boundary at $y=0$ between two different PhCs sharing the same crystal lattice, and exciting two counter-propagating edge states. Alternative configurations are described in Supplemental Material, including the excitation of guided modes of a linear defect inside a PhC. For simplicity, we focus on two-dimensional (2D) PhCs that do not rely on a photonic bandgap (PBG) for their confinement in the $z$ dimension, but most of the results can be generalized to 3D. We further assume that the PhCs are non-magnetic and lossless.\par

Because the structure is still periodic in the $x$ (albeit not in the $y$) direction, we choose an expanded "supercell" of the photonic structure comprised of one unit cell (of either PhC) in $x$ and infinitely many in $y$ direction. The supercell is used to compute the 1D Bloch states $\mathbf{E}_{k,n}(\mathbf{r}) = \mathbf{u}_{n}(k,\mathbf{r}) \cdot \exp(ikx)$, where the supercell's normalized $x$-periodic eigenmodes $\mathbf{u}$ are characterized by their band number $n$, wavenumber $k$ along the interface, and eigenfrequency $\omega_{n}(k)$. The eigenmodes can be sub-divided into two classes: (i) projected \cite{mold} bulk (extended) modes that have oscillatory behavior in $y$, and (ii) edge modes that exponentially decay as $e^{-\kappa |y|}$ away from the domain wall at $y=0$, where $\kappa^{-1}(\omega)$ is the localization distance. The focus of our calculation is on the edge modes that occupy all, or part, of the common bandgap of the two PhCs: $\omega_{\rm lb} < \omega < \omega_{\rm ub}$, where $\omega_{\rm l(u)b}$ are the lower (upper) bandgap edges.

The radiated electric field is calculated by solving the wave equation in the frequency domain: $\bigtriangledown\times(\bigtriangledown \times \mathbf{\tilde{E}}(\mathbf{r},\omega)) = (\omega/c)^2 \epsilon(\mathbf{r}) \mathbf{\tilde{E}}(\mathbf{r},\omega) + i\omega\mu_0 \mathbf{\tilde{J}}(\mathbf{r},\omega)$, where $\epsilon(\mathbf{r})$ represents the inhomogeneous dielectric permittivity of the entire structure, and $\mathbf{\tilde{J}}(\mathbf{r},\omega) = q\hat{r}_\parallel \delta^2(\hat{\mathbf{r}}_\perp) \exp(i\omega r_\parallel/v)$ is the current density produced by the charge moving with the constant speed $\mathbf{v} = v \hat{r}_\parallel$ in the direction of $\hat{r}_\parallel = \mathbf{v}/v$, $r_\parallel = \mathbf{r}\cdot\hat{r}_\parallel$, and $\mathbf{r}_\perp \perp \hat{r}_\parallel$ are the two remaining spatial dimensions.

In the case of a continuous medium on both sides of the boundary, the TR problem has been solved \cite{ginz,de2010optical} by stitching the analytically known solutions at the boundary. This approach is not workable in the case of PhCs because analytic solutions for the propagating waves cannot be obtained. However, the problem is simplified in the case of edge wave excitation due to the remaining periodicity in the $x$ direction. Briefly, using the Bloch eigenmodes of the supercell as the expansion basis \cite{sakoda}, the driven electric field can be expressed as an integral over the 1st Brillouin zone
\begin{equation}\label{main}
\mathbf{\tilde{E}}(\mathbf{r},\omega) = q \sum_n \int_{\mathrm{BZ}} \frac{\mathrm{d}k}{2\pi \epsilon_0} \ \frac{i  \omega c_{n}(k,\omega) \mathbf{E}_{k,n}(\mathbf{r})}{(\omega+i\gamma)^2 - \omega^{2}_{n}(k)} ,
\end{equation}
where the expansion coefficients $c_{n}(k,\omega)$ are given by an integral along the beam's path defined as $\mathbf{r} = r_\parallel\hat{r}_\parallel$:
\begin{equation}\label{overlap}
    c_n(k,\omega) =
    \int_{-\infty}^\infty \mathrm{d}r_\parallel (\mathbf{u}^*_{n}(k,\mathbf{r}) \cdot \hat{r}_\parallel) e^{i \left( \omega/v - k \cos{\theta} \right) r_\parallel},
\end{equation}
where $\theta$ is the angle between the directions of the beam's velocity and of the interface between the two PhCs. The summation over $n$ includes all modes (edge and bulk), and an infinitesimal $\gamma$ is introduced to ensure causality.\par

Only a discrete set of edge modes contributes to far-field radiation at frequencies inside the common bulk bandgap, thus enabling the following asymptotic limit of Eq.~(\ref{main}) (see Supplemental Material) at $x\to+\infty$:
\begin{equation}\label{EField}
    \mathbf{E}(\mathbf{r},t) \approx \sum_{m+} \int_{\omega_{\rm lb}}^{\omega_{\rm ub}} \frac{d\omega}{v^{(g)}_{m+}}  \frac{q C_{m+} \mathbf{u}_{m+}}{4\pi \epsilon_0} e^{i(k_{m+} x - \omega t)},
\end{equation}
where $m+$ is the discrete index for all forward-propagating edge modes, with their corresponding wave numbers \{$k_{m+}(\omega)$\} determined from the edge mode's dispersion relation $\omega_{m+}(k) = \omega$ and satisfying the causality condition $v^{(g)}_{m+}(\omega) \equiv \left( \mathrm{d}k_{m+}/\mathrm{d}\omega \right)^{-1} > 0$. The frequency-dependent spectral amplitudes $C_{m+}(\omega)$ of the transition edge radiation (TER) are obtained by substituting the implicitly frequency-dependent Bloch eigenfunctions $\mathbf{u}_{m+}(k_{m+},\mathbf{r})$ of the edge modes into Eq.(\ref{overlap}): $C_{m+} \equiv c_{m+}(k_{m+}(\omega),\omega)$. The expression for the electric field propagating in the $x<0$ direction is identical to Eq.(\ref{EField}), except that the contributing modes (labeled with $m-$ index) satisfy $v^{(g)}_{m-}(\omega) < 0$.

The power spectrum $P^{\rm TER}_{\pm}(\omega)$ of the forward/backward TER, which is finite for all frequencies where edge modes exist, can now be calculated (see Supplemental Material):
\begin{equation}\label{power}
    P^{\rm TER}_{\pm}(\omega) = \frac{q^2}{4\pi\epsilon_0} \sum_{m \pm} \frac{|C_{m \pm}|^2(\omega)}{v^{(g)}_{m \pm}(\omega)}
\end{equation}
This intuitive expression for the spectral power of edge waves, which is applicable to both continuous (see Supplemental Material for the application of this formalism to SPP generation \cite{ginz,de2010optical}) and photonic media, constitutes the main general result of this work.\par

\begin{figure*}
\includegraphics[width=\linewidth]{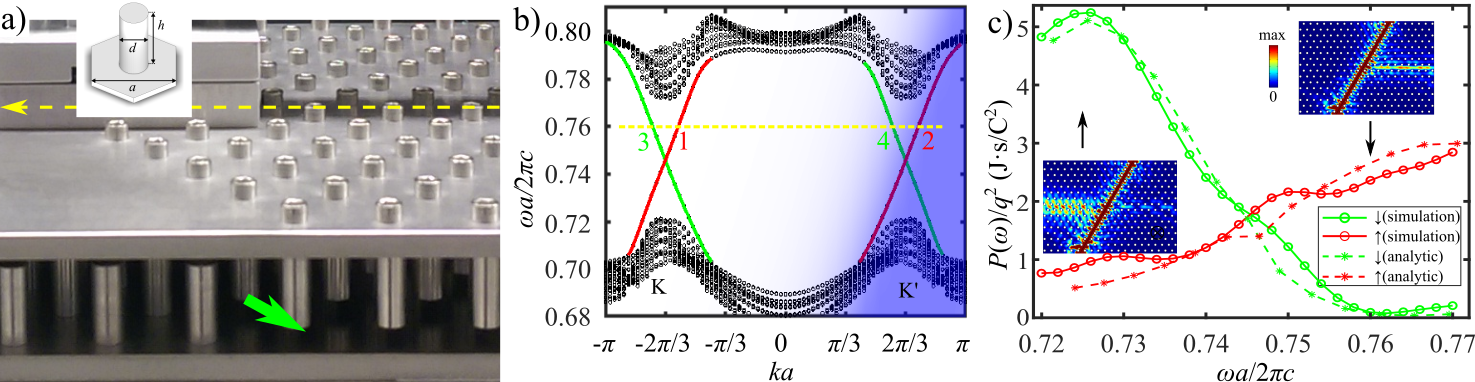}
\caption{\label{fig:PTI} (a) Fabricated photonic structure comprised of two QSH-PTIs separated by the domain wall. The charge travels in the mid-plane between the plates under the dashed yellow line. The spin-down waves (green arrow) will be received at one end of the interface (spin-up waves not shown). Inset: unit cell geometry with realistic dimensions: $a$=11.5 mm, $d$=3.97 mm, $h$=9.78 mm. (b) The 1D PBS of the structure, projected onto the $x$ axis. Black dots: bulk modes continua separated by the bandgap. Red/green solid lines: TPEWs with up/down ($m_{+}/m_{-}$) spins inside the bandgap. Horizontal dashed line: constant-frequency line intersecting the four TPEWs at different values of $k_{m\pm}(\omega)$. Blue shaded area: the "strong excitation belt". (c) The emitted TER inside the bandgap calculated from the analytic expression (Eq.~(\ref{power}), dashed lines) and \emph{ab initio} simulation (solid lines). Red/green lines: TER to the right/left of the crossing point. Insets: the norm of the in-plane Poynting vector perpendicular to the beam's path $|S_\perp|$ obtained from the simulation, at frequencies labeled by the black arrows. The beam moves form the lower to the upper PTI domain, and its trajectory is covered by over-saturated red. The horizontal PTI interface is located in the middle of the plot, below/above which rods are attached to the top/bottom plate.}
\end{figure*}

Next, we consider a specific example of kink states' excitation at the domain wall between two topologically-different PTIs shown in Fig.~\ref{fig:PTI}(a). The structure, based on Ref.~\cite{PhysRevLett.114.127401}, consists of two quantum spin-Hall (QSH) PTIs with opposite spin Chern numbers $C_s=\pm 1/2$. The QSH-PTIs are comprised of two parallel metal plates providing confinement in the $z$ direction, patterned by a hexagonal lattice (of period $a$) arrangement of metal rods attached to either the top (right side of Fig.~\ref{fig:PTI}(a)) or the bottom (left side) metal plate. Its 1D-photonic band structure (PBS) and Bloch states are obtained using \textsc{comsol} eigenfrequency study. Fig.~\ref{fig:PTI}(b) shows the PBS, where black dots denote bulk modes, and colored solid lines inside the bandgap represent TPEWs.

The domain wall between two QSH-PTIs supports four TPEWs inside the bandgap: two forward and two backward TPEWs (two at each valley). The group velocities of the TPEWs are locked to their photonic spin \cite{PhysRevLett.114.127401}: spin-up ($m+$, red lines) modes propagate forward, while spin-down ($m-$, green lines) modes propagate backwards. For our specific design, the TPEWs span the shared topological bandgap bracketed by $\omega_{\rm lb} = 0.72(2\pi c/a)$ and $\omega_{\rm ub} = 0.77(2\pi c/a)$ from below and above, respectively. The TER-producing point charge is assumed to be moving along one of the high-symmetry axes of the hexagonal lattice, drawn through the mid-plane between the two metal plates and half-way between two adjacent rows of rods for optimal clearance. Therefore, the charge is crossing the domain wall between the two QSH-PTIs at the $\theta=\pi/3$ angle (see Fig.~\ref{fig:PTI}(a)), and is experiencing a periodic environment on both sides of the interface.

The choice of this specific photonic platform is dictated by its several unique properties. First, the supported TPEWs can be guided along sharply curved trajectories \cite{PhysRevLett.114.127401,SciRep,khanikaev2013photonic,chen2014experimental} after their excitation. Second, the specific geometry of QSH-PTIs is conducive to its interaction with high-power electromagnetic radiation. That is because the transverse confinement of the kink states does not require any side walls, and because the attachment of the rods to just one metal plate enables their easy monolithic fabrication. Third, the sparsity of the QSH-PTI structure and the existence of clear passages for the charged beam along multiple unobstructed directions prevent a direct impact of electrons on the structure. We note that it has been recently shown in theory that unidirectional edge states can be predominantly excited by Cherenkov emission using magnetized plasmas or Weyl semi-metals \cite{PhysRevB.98.115136}.

The expression for the power spectrum $P^{\rm TER}_{\pm}(\omega)$ involves $4$ TPEWs that are graphically shown as the crossing points between the yellow dashed (constant frequency) line and the dispersion relations (solid lines) of the TPEWs in Fig.~\ref{fig:PTI}(b). These crossings are labeled as follows:  $m=1,2$ crossings belong to $\{m+\}$ (spin-up TPEWs in the $K/K^{\prime}$ projected valleys), while $m=3,4$ correspond to their spin-down counterparts. The group velocities of all $4$ TPEWs are approximately equal and constant across the bandgap: $v^{(g)}_{m}(\omega) \approx 0.4c$. The predicted spectra are plotted in Fig.~\ref{fig:PTI}(c) for the right/left-propagating TPEWs (dashed red/green lines), and are found in good agreement with \emph{ab initio} driven simulation (solid lines), where $\mathbf{\tilde{J}}(\mathbf{r},\omega)$ is implemented as the current source.

The TER spectra exhibit several notable features. First, we find that TER can be highly directional and spin-polarized: see the insets in Fig.~\ref{fig:PTI}(c) corresponding to $\omega_{\downarrow} \approx 0.725 (2\pi c/a)$ (predominantly backward spin-down radiation), and to $\omega_{\uparrow} \approx 0.76 (2\pi c/a)$ (forward spin-up radiation). On the other hand, for other frequencies at the center of the bandgap both forward and backward TPEWs of similar intensities are launched. Second, excitation of $K$ valley TPEWs ($m=1,3$) is negligible compared with  excitation of their $K'$ valley ($m=2,4$) counterparts (see Fig.~S2 in Supplemental Material for the spectra of all $4$ TPEWs). Therefore, transition radiation mechanism provides a new way of valley-polarized excitation of TPEWs, and provides an opportunity to introduce the concept of quasi-phase matching (QPM) between charges and radiation.

The essence of QPM is that under the envelope function approximation \cite{bastard1988wave}, TPEWs are constructed from bulk modes of the 2D-periodic PhC with imaginary $k_y^{\rm bulk}$ (for the QSH-PTI we used, the edge mode with 1D wavevector $k$ is constructed from bulk modes with purely imaginary $k_y^{\rm bulk}=\pm i\kappa\approx\pm0.38ia^{-1}$ and real $k_x^{\rm bulk}= k\pm2\pi/a$ for $K(k<0)$ and $K'(k>0)$ valley, respectively, due to band folding \cite{ma2016all}), and for weakly confined-TPEWs, the projection of the real part of the 2D wavevector $\mathbf{k}^{\rm bulk}$ onto the charge trajectory must approximately match the wavenumber $\omega/v$ of the line current (or differ by a reciprocal vector) in order to get large overlap integral Eq. (\ref{overlap}). Quantitatively, strong excitation of an edge mode is possible when its $\omega$ and $k_x^{\rm bulk}$ satisfies
\begin{equation}\label{eq:SER}
  |\omega/v-k_x^{\rm bulk}\cos\theta+2\pi N/a'|\lesssim\kappa\sin\theta,
\end{equation}
for some integer $N$, where $a'$ is the period along the beam's path. Note that in the limit of $\kappa\sin\theta\to0$, we recover the so-called generalized Cherenkov condition \cite{Kroha}. We refer to the region defined by Eq. (\ref{eq:SER}) as "strong excitation belt", and it is graphically represented in Fig.~\ref{fig:PTI}(b) as the blue shaded area ($N=-1$ for $K'$ valley and $a'=a$).
It's clear that only TPEWs at $K'$ valley fall into this belt, and this explains why they are predominantly excited. One can also see the reason why the backward-moving TPEW $4$ is excited much stronger at the lower edge of the bandgap than at the upper edge: its dispersion line lies deep inside the belt at lower frequencies but outside at higher frequencies. Additional examples corresponding to a sub-relativistic beam with $v=0.56c$ (strong excitation belt covering $K$ valley modes) and $v=0.75c$ (no excitation), as well as detailed derivation of Eq. (\ref{eq:SER}) are presented in Supplemental Material. The total emitted energy in the topological bandgap versus $v$ is also plotted, where multiple sharp peaks correspond to significantly enhanced TER when the QPM condition is satisfied. QPM is important when (i) the edge mode can be well described by decaying bulk modes and (ii) the wave decay constant $\kappa$ satisfies $\kappa a'\sin\theta\ll2\pi$. Note that, formally, the QPM condition resembles the relationship between the emission angle $\theta$ and the frequency $\omega$ of the Smith-Purcell radiation produced by a charge moving along a periodic structure \cite{yamamoto2015interference,kaminer2017spectrally}. The key differences in the case of TER considered by us are as follows: (i) radiation is coupled into a discrete set of edge states, not into a bulk continuum, (ii) the emission angle $\theta$ is fixed by the relative orientations of the charge trajectory and the interface, and (iii) due to the transient nature of TER, the QPM condition is an inequality rather than a strict equation.

\begin{figure}
\includegraphics[width=\linewidth]{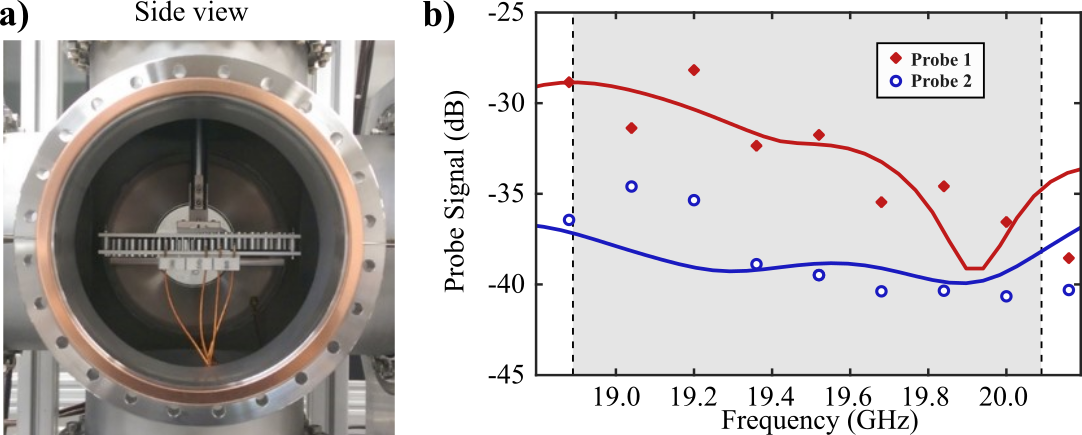}
\caption{\label{fig:exp} Experimental demonstration of transition edge-wave radiation (TER) using  two interfaced QSH-PTIs. (a) fabricated structure inside the chamber. (b) Experimentally measured signals by Probe 1 (red diamonds) and 2 (blue circles). Numerical prediction: solid red (blue) curves for Probe 1 (2). Shadowed area: photonic bandgap, where only TPEWs exist.}
\end{figure}

The experimental validation of the TER concept was carried out at the Argonne Wakefield Accelerator Facility (AWA-ANL) using a high-charge relativistic point-like electron beam ($q \sim 3$ nC, $E_b \approx 65$ Mev, and $\tau_b \approx 3$ ps) and a photonic structure that was modeled above. As sketched in Fig. \ref{fig:PTI}(a), the bunch (yellow dashed line) traverses the interface between the two QSH-PTI domains near the center of the structure and excites TPEWs around the frequency of $f_0 \equiv \omega_0/2\pi  \approx 19.5$ GHz. The fully-assembled structure is pictured in Fig. \ref{fig:exp}(a) inside a vacuum chamber. The objectives were to experimentally demonstrate the following: (a) unobstructed charged bunch propagation over many periods of the PhC along one of its principal directions under full vacuum; (b) the capability of the TR mechanism to exciting TPEWs inside the bulk bandgap, and (c) spatial localization of TPEWs close to the domain wall.  The PhC was comprised of $15\times 13$ unit cells, and its dimensions listed in the caption of Fig.~\ref{fig:PTI} ensure a topological PBG in the $19 < f < 20$ GHz range (where $f =\omega /2\pi$).

For diagnosing the TER produced by the bunch, two probes were positioned along the outer edge of the structure to detect spin-down waves: one very close (Probe $1$), the other (Probe $2$) 6 periods away from the interface (see Supplemental Material for their exact positions). The comparison between the signals from the two probes (see Fig. \ref{fig:exp}(b)) is used to demonstrate spatial localization of the EM energy at the interface. Indeed, the measured signal from Probe 1 (red diamonds) is much stronger that that from Probe 2 (blue circles) for every frequency inside the PBG (shadowed area). This contrast, which approaches $10$ dB for some of the frequencies, implies that the beam excites edge waves. Topological protection of such modes has been demonstrated earlier \cite{SciRep} using antenna excitation. To our knowledge, this is the first time that TPEWs were shown to be excited via the TR mechanism. Due to limitation of the present experimental setup, we were only able to reliably measure signals at one end of the interface. Nonetheless, the measured result captures the most salient features predicted by our theory. For example, we see a clear trend of the TER decreasing in power as the frequency increases from the lower to the upper edge of the PBG. This is a consequence of the breakdown of the QPM near the upper edge of the PBG, as predicted by numerical simulation (solid red curve) and discussed above. Although the CR produced by the beam outside of the bandgap is beyond the scope of this Letter, we note that the frequency positions (at 16GHz and 18 GHz) of its two measured spectral peaks are also in good agreement with simulations results. Additional experimental and data processing details can be found in Supplemental Material. Future improvement of the experiment includes measuring spin-down waves and using a long train of electron bunches.

The TER concept can be used for beam diagnostics in the same way as TR of bulk waves because $P^{\rm TER}_{\pm}(\omega)$ strongly depends on the beam's energy, duration, and the location of its trajectory. Novel beam-driven accelerators, such as matrix \cite{matrix} and two-beam accelerators (TBAs) \cite{kazakov2010high}, and accelerators with a photonic-band-gap structure \cite{smirnova2005demonstration} can also benefit from TER. For example, possible geometries of a TER-based non-collinear (but parallel) TBA and a matrix accelerator are shown in Supplemental Material.\par

\begin{acknowledgments}
This work was supported by the Army Research Office (ARO) under Grant No. W911NF-16-1-0319, and by the U.S. Department of Energy (DOE) Office of Science under Grant  No. DE-SC0007889.
\end{acknowledgments}
\bibliography{TR_Refs}

\end{document}


\maketitle
\section{Some other configurations using TR to excite guided modes}
\begin{figure}[h!]
\centering
\includegraphics[width=\linewidth]{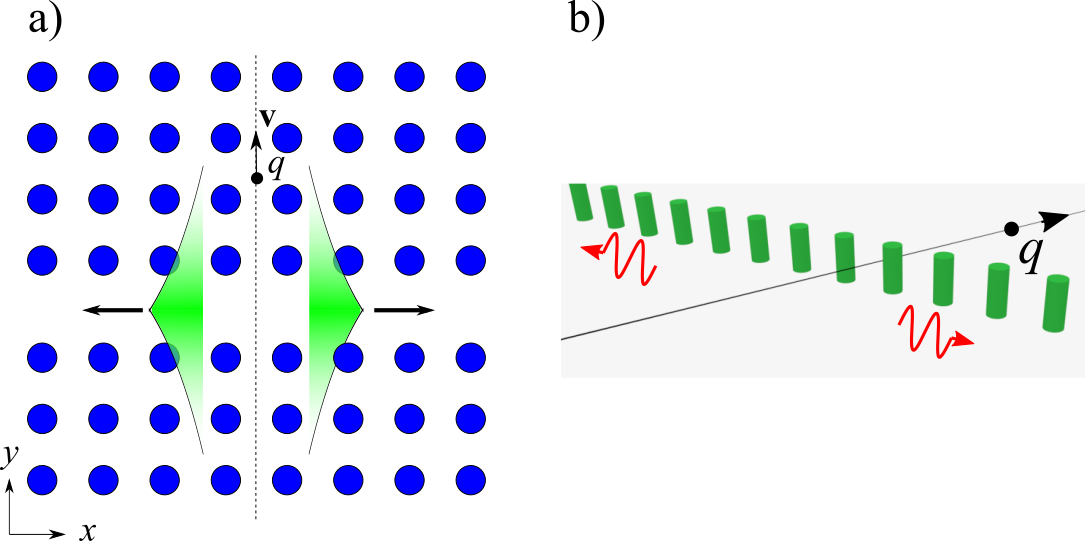}
\caption{Some other configurations using TR to excite guided modes. (a) a charge crossing a linear defect in a PhC can excite the waveguide modes localized at the defect. (b) a charge moving closely past an array of rods forming a "periodic dielectric waveguide" will also excite guided modes.}
\label{fig:waveguide}
\end{figure}

As mentioned in the main text, the formalism of guided waves excitation by transition radiation we present is quite general and can be applied to many configurations other than the one shown in the Fig. 1, where kink modes between two different PhCs are excited. One common example is shown in Fig.~\ref{fig:waveguide}(a),
where a charge crossing a linear defect (for example, removing a row of rods) in a PhC can excite the waveguide modes localized at the defect. Another interesting example is shown in Fig.~\ref{fig:waveguide}(b), where a charge moving closely past an array of dielectric rods, which forms a so-called "periodic dielectric waveguide" \cite{mold}, will excite these guided modes. This configuration is interesting because here the difference between TR and diffraction radiation becomes fussy, since in a PhC it's hard to tell whether the charge actually penetrates the structure.
\section{Asymptotic behavior of electric field}
First, for the sake of completeness of derivation, we mention here how the eigenmodes are chosen. The eigenmodes used to perform the expansion satisfy the homogeneous wave equation (i.e., where $\mathbf{J}=0$) and the following orthogonality/normalization conditions:
\begin{equation}\label{norm}
    \int\mathrm{d}^3r\epsilon(\mathbf{r})\mathbf{E}_{k,n}^*(\mathbf{r})\cdot\mathbf{E}_{k',n'}(\mathbf{r})=2\pi\delta_{nn'}\delta(k-k').
\end{equation}\par
Now we derive the asymptotic behavior of electric field far away from the source (where the charge cross the boundary): in Eq. (1), when $|x|/a\to\infty$ but $y$ remains finite, both $c_{k,n}$ and $\mathbf{u}_{k,n}(\mathbf{r})$ are mathematically well-behaved. If $x>0$, we can close the contour in Eq. (1) in the upper-half of the complex $k$-plane, and use the fact that the integral along the semicircle vanishes as $x\to\infty$ by Jordan's lemma. Using the residue theorem, we obtain:
\begin{align}
    \int_{-\pi/a}^{\pi/a}&\frac{\mathrm{d}ke^{ikx}c_{k,n}\mathbf{u}_{k,n}(\mathbf{r})}{(\omega-\omega_{k,n}+i\gamma)(\omega+\omega_{k,n}+i\gamma)}
    \nonumber\\
    =&2\pi i\sum_ {m+}\left.\frac{c_{k,n}\mathbf{E}_{k,n}(\mathbf{r})}{-2\omega\cdot\mathrm{d}\omega/\mathrm{d}k}\right|_{k=k_{m+}},
\end{align}
where, for the $\omega$ residing in one of the common bandgaps, \{$k_{m+}$\} are solutions of equation $\omega_{k,n}=\omega$ that simultaneously satisfy the causality condition $v_g :=\mathrm{d}\omega/\mathrm{d}k>0$ (i.e. energy propagates from the $x=0$ point where the beam crosses the boundary to $x>0$). Naturally, the causality requirement for $x<0$ is $\mathrm{d}\omega/\mathrm{d}k<0$.
\section{Some useful identities of Bloch eigenmodes}\label{identity}
In this sections we prove two useful identities
of Bloch eigenmodes, which will be used in the next section.\par
First, in a lossless media, modes associated with different $k$, $(\mathbf{E}_1,\mathbf{H}_1)$ and $(\mathbf{E}_2,\mathbf{H}_2)$ are orthogonal in the sense that $\int\mathrm{d}\mathbf{S}\cdot \Re[\mathbf{E}_1\times \mathbf{H}_2^*+\mathbf{E}_2\times \mathbf{H}_1^*]=0$. To show this, note that since they are solutions of the homogeneous Maxwell solution, we have
\begin{align}
    0=&\mathbf{E}_1\cdot(\bigtriangledown\times\mathbf{H}_2^*-i\omega\mathbf{D}_2^*)\nonumber\\
    =&-\bigtriangledown\cdot(\mathbf{E}_1\times\mathbf{H}_2^*)-i\omega(\mathbf{E}_1\cdot\mathbf{D}_2^*-\mathbf{B}_1\cdot\mathbf{H}_2^*).
\end{align}
Therefore
\begin{align}
    &\bigtriangledown\cdot(\mathbf{E}_1\times\mathbf{H}_2^*+\mathbf{E}_2\times \mathbf{H}_1^*)\nonumber\\
    =&-i\omega(\mathbf{E}_1\cdot\mathbf{D}_2^*-\mathbf{B}_1\cdot\mathbf{H}_2^*+\mathbf{E}_2\cdot\mathbf{D}_1^*-\mathbf{B}_2\cdot\mathbf{H}_1^*).
\end{align}
In a lossless media, the last line is purely imaginary, so $\Re[\bigtriangledown\cdot(\mathbf{E}_1\times\mathbf{H}_2^*+\mathbf{E}_2\times \mathbf{H}_1^*)]=0.$ Now integrate this term in a unit cell and convert it to surface integrals
\begin{align}\label{orth}
    0=&\int_{x=x_0}\mathrm{d}\mathbf{S}\cdot\Re[\mathbf{E}_1\times \mathbf{H}_2^*+\mathbf{E}_2\times \mathbf{H}_1^*]-\int_{x=x_0+a}\mathrm{d}\mathbf{S}\cdot\Re[\mathbf{E}_1\times \mathbf{H}_2^*+\mathbf{E}_2\times \mathbf{H}_1^*]\nonumber\\
    =&(1-e^{i(k_1-k_2)a})\int_{x=x_0}\mathrm{d}\mathbf{S}\cdot\Re[\mathbf{E}_1\times \mathbf{H}_2^*+\mathbf{E}_2\times \mathbf{H}_1^*],
\end{align}
where we have use the fact that the eigenmodes are edge modes so there is no power outflow in the $y$ direction. Since $k_1\neq k_2$ and they are restricted in the first Brillouin zone, the prefactor is non-zero. This proves the orthogonal relation we want.\par
Second, if we use only one mode in Eq. (\ref{orth}) instead of two different ones, then the first term vanishes because of periodicity, and we immediately see that electric energy equals the magnetic energy for any eigenmodes.
\section{Derivation of the expression of energy spectral density}
Since the total energy $U$ emitted in the process of transition radiation is finite, we can use the energy spectral density $P(\omega)$, defined as
\begin{equation}
    U=\int\mathrm{d}\omega P(\omega),
\end{equation}
to describe its spectral distribution. As discussed above, we only discuss frequencies in the bulk bandgap.\par
It can be easily shown (see Appendix A of Ref.~\cite{Kroha} for a similar proof) that
\begin{equation}
 P(\omega)=\frac{2}{\pi}\int\mathrm{d}\mathbf{S}\cdot\frac{1}{2}\Re[\mathbf{E}(\mathbf{r},\omega)\times\mathbf{H}^*(\mathbf{r},\omega)],
\end{equation}
where the integral surface $S$ is the $yz$ plane perpendicular to the direction of the energy flow, and located at large values of $x$ (i.e. far from the crossing point).\par
Using the "orthogonal" property we proved in the last section, we see all the cross terms vanish, and the energy spectral density in $+x$ direction is
\begin{align}\label{power0}
    P(\omega)
    =&\frac{2q^2}{\pi(2\epsilon_0)^2}\sum_{m+}\left.\frac{|c_{k}|^2P_{k}}{v_g^2}\right|_{k=k_{m+}},
\end{align}
where $P_{k}$ is the power flow of a single mode calculated as
\begin{equation}
    P_{k}=\int\mathrm{d}\mathbf{S}\cdot\frac{1}{2} \Re[\mathbf{E}_k\times\mathbf{H}^*_k].\end{equation}\par
Using the fact we just proved that electric energy equals the magnetic energy for any eigenmodes, and that the group velocity equals the energy transport velocity for a lossless mode \cite{mold}, we further simplify the previous equation:
\begin{align}\label{energy velocity}
P_{k}=\frac{1}{4}\epsilon_0v_g\times2,
\end{align}
since the normalization condition in this case is
\begin{align}
    \int\mathrm{d}y\int\mathrm{d}z\int_0^a\mathrm{d}x\epsilon(\mathbf{r})|\mathbf{E}_k(\mathbf{r})|^2=a.
\end{align}
Substituting Eq.(\ref{energy velocity}) into Eq.(\ref{power0}) yields
\begin{align}
    P(\omega)=\frac{q^2}{4\pi\epsilon_0}\sum_{m+}\left.\frac{|c_{k}|^2}{v_g}\right|_{k=k_{m+}}.
\end{align}
Changing the label from $m+$ to $m-$ expresses the energy spectral density in the $-x$ direction.

\section{Emission of surface plasma polaritons by transition radiation}\label{spp}
When a charge cross the boundary of two media with different $\epsilon$, it will excite surface modes, called surface plasma polaritons (SPPs). This is discussed in detail in Ginzburg's book \cite{ginz}, although the terminology of SPP is not used. Since our approach, based on the assumption of PhCs, is more general, it should give the same answer. We will demonstrate it here.\par
We assume that the metal is at $z<0$, and dielectric at $z>0$. The normalized eigenmodes are
\begin{align}
\mathbf{E_{\perp,\mathbf{k}}}&=A\exp(\pm z\sqrt{k^2-\epsilon\omega_k^2/c^2})\hat{\mathbf{k}},\\
E_{z,k}&=\pm \frac{k}{\sqrt{k^2-\epsilon\omega_k^2/c^2}}A\exp(\pm z\sqrt{k^2-\epsilon\omega_k^2/c^2}),\label{1}
\end{align}
where
\begin{align}
A^2&=\frac{2k|\epsilon_1\epsilon_2|^{1/2}}{|\epsilon_2^2-\epsilon_1^2|},\\
\omega_k^2&=\frac{(\epsilon_1+\epsilon_2)k^2 c^2}{\epsilon_1\epsilon_2}.
\end{align}
The plus sign in Eq. (\ref{1}) is for the metal, which we will use subscript 1 to denote afterwards, and minus sign and subscript 2 for the dielectric.\par
Using the result (Eqs. (1, 2)) from the main manuscript, the Fourier component of the $E_z$ field in the metal is
\begin{align}
E_{z,k,1}(\omega)=&\frac{-i}{(2\pi)^2}\frac{\omega}{\epsilon_0}\frac{E_{z,k,1}}{\omega^2-\omega_k^2}(\int_{-\infty}^0\mathrm{d}z'E_{z,k,1}qe^{i\omega z'/v}+\nonumber\\
&\int_0^{\infty}\mathrm{d}z'E_{z,k,2}qe^{i\omega z'/v})\nonumber\\
=&\frac{-iq}{(2\pi)^2}\frac{\omega}{\epsilon_0}\frac{1}{\omega^2-\omega_k^2}\frac{2k|\epsilon_1\epsilon_2|^{1/2}}{|\epsilon_2^2-\epsilon_1^2|}\epsilon_2\times\nonumber\\
&\exp(z\sqrt{k^2-\epsilon_1\omega_k^2/c^2})\times\nonumber\\
&(\frac{1}{\epsilon_1}\frac{1}{\sqrt{k^2-\epsilon_1\omega_k^2/c^2}+i\omega/v}+\nonumber\\
&\frac{1}{\epsilon_2}\frac{1}{\sqrt{k^2-\epsilon_2\omega_k^2/c^2}-i\omega/v})
\end{align}
This result is not exactly the same to (2.20) and (2.28) in Ref.~\cite{ginz}, but this is because here in the eigenmode expansion, we only take into account the surface mode. To get the exact field, one need to include bulk modes (extended modes in either media) as well, which actually form a continuum in the 1D dispersion diagram. But if we are mainly interested in surface waves ($ \mathbf{r}_\perp\to\infty$ but $z$ finite), in which case we should take $\omega\to\omega_\kappa$ (the so-called plasmon-pole approximation \cite{ford1984electromagnetic}), then the result becomes identical to that in Ref.~\cite{ginz}.\par

\section{Formation Length in TER}
One important concept in TR is the length $L_f$ of the formation zone, based on the fact that radiation is not formed at a point but within a finite region \cite{ginz}. In bulk TR, it is usually defined as the distance where the interference of the charge field and the radiation field is still important, and therefore determined by the phase difference of the two fields along the beam's path. Here for TER in PhCs, for simplicity, we discuss the case when $\theta=\pi$ (normal incident). The envelope of the charge field will have a phase dependence of $\omega y/v$; while the radiation field does not have phase variance along $y$ direction, it exponentially decays away from the interface as $e^{-\kappa |y|}$. The decay rate should play a role in the formation length $L_f$ since if the edge mode is very localized, the charge can only interact with it in a short distance. Therefore it is reasonable to include both the phase difference and decay rate and define $L_f\sim2\pi|\omega/v+i\kappa|^{-1}$ as the formation length.

\section{Quasi-phase matching between edge modes and the current}\label{matching}
The QSH-PTI used in the main text has a Kane-Mele Hamiltonian \cite{PhysRevLett.114.127401}. By performing an calculation similar to the one in Ref.~\cite{wada2011localized}, one can verify that edge modes only exist when the two sides have opposite signs of $\Delta_{em}$ (using quantities defined in Ref.~\cite{PhysRevLett.114.127401}, same below) and thus opposite spin-Chern numbers, get the edge mode dispersion $\Omega=v_D\delta k_x$ ($\Omega$ and $\delta k_x$ are detuning frequency and wavevector from the Dirac point), and see that the edge modes with 1D wavevector $k$ are constructed from bulk modes with with purely imaginary $k_y^{\rm bulk}=\pm i\kappa$, $\kappa=\Delta_{em}\omega_D/v_D\approx0.38a^{-1}$ and real $k_x^{\rm bulk}= k\pm2\pi/a$ for $K(k<0)$ and $K'(k>0)$ valley, respectively. The additional term $\pm2\pi/a$ is due to band folding when going from 2D Brillouin zone to 1D \cite{ma2016all}. So the electric field of edge modes can be written as (subscript $n$ is not written out for conciseness)
\begin{equation}
    \mathbf{E}_k(\mathbf{r})=\mathbf{u'}_k(\mathbf{r})e^{ik_x^{\rm bulk}x}e^{-\kappa|y|}.
\end{equation}
Here $\mathbf{u'}(\mathbf{r})$ is a periodic function on a unit cell.
Using this expression, and note that $y=r_\parallel\sin\theta$, Eq.~(2) would become
\begin{align}\label{eq:overlap}
    c(k,\omega)=&\int_{-\infty}^\infty\mathrm{d}r_\parallel(\mathbf{u'}^*_{k}(\mathbf{r})\cdot\hat{r}_\parallel)e^{i(\omega/v-k_x^{\rm bulk}\cos\theta)r_\parallel-\kappa\sin\theta |r_\parallel|}.
\end{align}
Now that $\mathbf{u'}^*_{k}\cdot\hat{r}_\parallel$ is a periodic function with periodicity $a'$, we can expend it as a Fourier series: $\mathbf{u'}^*_{k}\cdot\hat{r}_\parallel=\sum_{N=-\infty}^{\infty}c_N(k)e^{2\pi iNr_\parallel/a'}$. Then the spectral amplitude (subscript $m\pm$ is not written out for conciseness)
\begin{align}
    C(\omega)\equiv&c(k(\omega),\omega)=\sum_{N=-\infty}^{\infty}c_N(k(\omega))\int_{-\infty}^\infty\mathrm{d}r_\parallel e^{i(2\pi N/a'+\omega/v-k_x^{\rm bulk}(\omega)\cos\theta)r_\parallel-\kappa\sin\theta |r_\parallel|}\nonumber\\
    =&\sum_{N=-\infty}^{\infty}c_N(k(\omega))\int_{0}^\infty\mathrm{d}r_\parallel\{e^{[i(2\pi N/a'+\omega/v-k_x^{\rm bulk}(\omega)\cos\theta)-\kappa\sin\theta]r_\parallel}\nonumber\\
    &+e^{[-i(2\pi N/a'+\omega/v-k_x^{\rm bulk}(\omega)\cos\theta)-\kappa\sin\theta]r_\parallel}\}\nonumber\\
    =&\sum_{N=-\infty}^{\infty}c_N(k(\omega))[\frac{1}{-i(2\pi N/a'+\omega/v-k_x^{\rm bulk}(\omega)\cos\theta)+\kappa\sin\theta}\nonumber\\
    &+\frac{1}{i(2\pi N/a'+\omega/v-k_x^{\rm bulk}(\omega)\cos\theta)+\kappa\sin\theta}]\nonumber\\
    =&\sum_{N=-\infty}^{\infty}c_N(k(\omega))\frac{2\kappa\sin\theta}{(2\pi N/a'+\omega/v-k_x^{\rm bulk}(\omega)\cos\theta)^2+(\kappa\sin\theta)^2}.
\end{align}
is the sum of a series of Lorentzian-like functions, assuming the Fourier coefficients $c_N(k)$ are slow varying functions of $k$, since $k(\omega)$ has a approximately linear dependence on $\omega$. If $\kappa a'\sin\theta \ll 2\pi$, the Lorentzian peaks will be well separated, and we can take the half width at half maximum as a criterion for strong excitation:
\begin{equation}
    |\omega/v-k_x^{\rm bulk}(\omega)\cos\theta+2\pi N/a'|\lesssim\kappa\sin\theta,
\end{equation}
or, if we define $k_b$ as the value of $k$ that lets the left hand side of the above inequality be zero for some integer $N$, this strong excitation condition can be expressed as
\begin{equation}\label{eq:excitation belt}
    |k(\omega)-k_b|\lesssim\kappa\tan\theta.
\end{equation}

Using the relation between $k$ and $k_x^{\rm bulk}$, we obtain the following formula for $k_b$ for the $K-$ ($k<0$) and $K^{\prime}-$valley ($k>0$) states, with $\theta=\pi/3$ and $a'=a$:
\begin{align}
  \omega &= \frac{k_b v}{2} + \frac{2\pi c}{a} \left( N - \frac{1}{2} \right) \frac{v}{c} \ \ \ \  {\rm for} \ \ \ \   k > 0 \label{eq:kx>0}\\
  \omega &= \frac{k_b v}{2} + \frac{2\pi c}{a} \left( N + \frac{1}{2} \right) \frac{v}{c} \ \ \ \  {\rm for} \ \ \ \  k < 0 \label{eq:kx<0},
\end{align}
where both $\omega$ and $v$ are assumed positive. Here, because of the specific choice of $\theta=\pi/3$, both valleys share the same series of $k_b$'s, but the value of $N$ depends on which valley is referred to. These lines can be referred to as "beam lines".\par

\begin{figure}[ht]
\centering
\includegraphics[width=\linewidth]{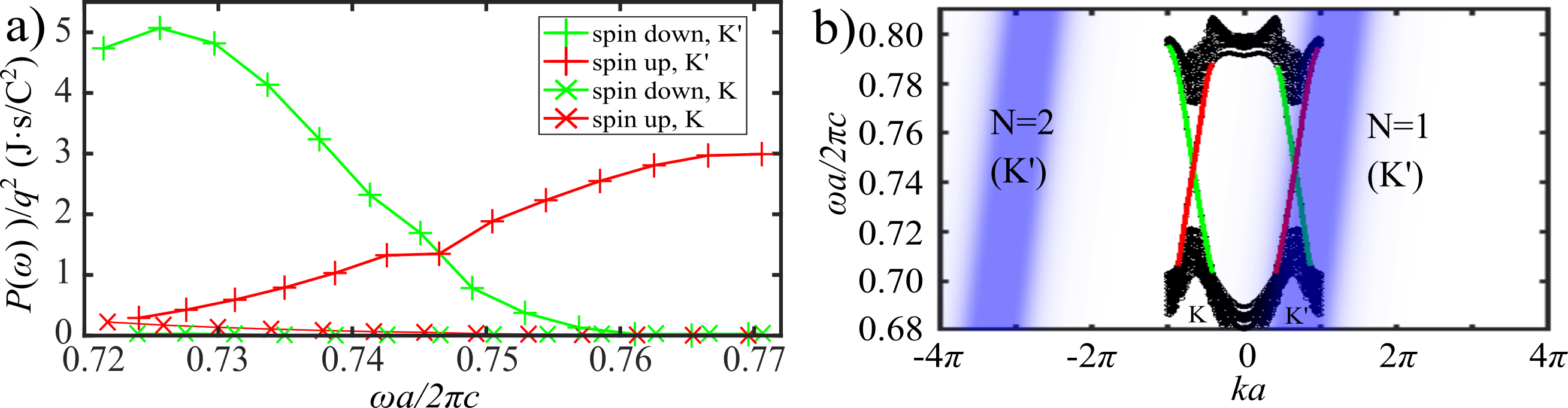}
\caption{(a) The spin- and valley- specific energy spectral density with a charge of $v\approx c$, where it is clear that edge modes at K' valley is dominantly excited. (b) The 1D PBS of the supercell and the "strong excitation belts" in blue corresponding to Eq. (\ref{eq:kx>0}) with $N=1,2$.}
\label{fig:S1}
\end{figure}

The inequality Eq. (\ref{eq:excitation belt}) will define a series of belts in the 1D PBS, which we shall refer to as "strong radiation belts". For $v=c$, they are shown as the blue belts in Fig. \ref{fig:S1}(b), corresponding to $N=1,2$ in Eq. (\ref{eq:kx>0}). Note that although the 1D PBS has a period of $2\pi/a$ and can thus be confined in the first Brillouin zone $-\pi/a<k<\pi/a$, the "strong excitation belts" do not (here they have a period of $4\pi/a$ instead), and $k_b$ can go far beyond the first Brillouin zone. The underlying reason is that $k_b$ is originated from the 2D PBS.\par

We note that the concept of the beam line introduced here to describe the most efficient excitation of the TR is fundamentally different from the much more familiar related concept of the $\omega_{\rm b}^{\rm res} = \mathbf{k} \cdot \mathbf{v}$ dispersion curve used in the context of the resonant (Cherenkov) radiation of electromagnetic waves. For example, in the context of Cherenkov excitation of TPEWs, one could use a beam co-propagating with the interface (e.g., in the $x-$direction), where the appropriate resonant beam dispersion can be expressed as $k = k_{\rm b}^{\rm res}(\omega) \equiv \omega/v$. Cherenkov radiation is excited at, and only at the frequencies $\omega^{\rm res}$ such that the following equation is satisfied: $k_{\rm b}^{\rm res}(\omega^{\rm res}) = k_n^{\rm TPEW}(\omega^{\rm res}) + 2\pi N/a$, where $N$ is an integer. Applications of such resonant wave-particle interactions to high power generation has been considered in the context of beam-powered traveling-wave microwave tubes~\cite{tsimring,goebel_ieee94} and, more recently, metamaterials and metawaveguides~\cite{shvets_shapiro_prb12}.

On the contrary, the transition radiation is not a resonant process. That's why "strong excitation belt" is a more appropriate condition for strong excitation than the "beam line". The condition Eq. (\ref{eq:excitation belt}) can be interpreted as "quasi-phase matching": although the edge modes decay away from the interface, they can still be assigned a "quasi-wave vector" along the charge trajectory. If it approximately matches the wave vector $\omega/c$ of the line current, the overlap integral in Eq. (\ref{eq:overlap}) becomes large, and the TPEW is strongly excited.\par

\begin{figure}[ht]
\centering
\includegraphics[width=\linewidth]{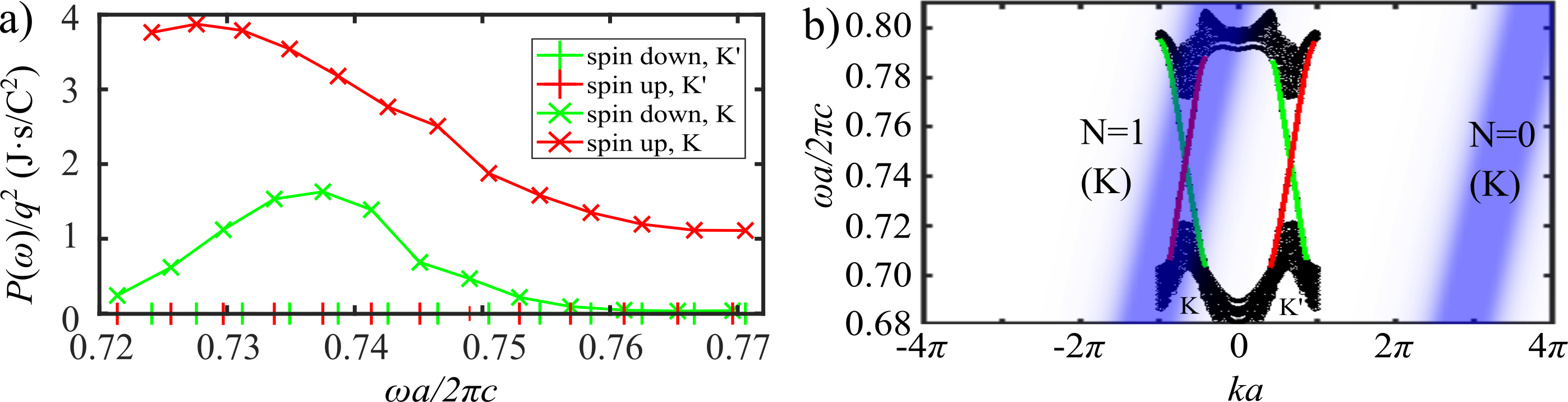}
\caption{(a) The spin- and valley- specific energy spectral density with a charge of $v=0.56c$, where it is clear that edge modes at K valley is dominantly excited. (b) The 1D PBS of the supercell and the "strong excitation belts" in blue corresponding to Eq. (\ref{eq:kx<0}) with $N=1,0$.}
\label{fig:S2}
\end{figure}

It is straightforward to apply this "quasi-phase matching" idea if we want to predominantly excite edge waves at the other valley. Simple calculation shows that changing $v$ to $0.56c$ while keeping the trajectory satisfies Eq.~(\ref{eq:kx<0}) with $N=1$, and Eq.~(\ref{eq:kx>0}) with $N$ far from any integer. The resultant energy spectral density shown in Fig.~\ref{fig:S2}(a) verifies our expectation, where both spin modes at K-valley is dominantly excited. The corresponding strong excitation belts are shown in Fig.~\ref{fig:S2}(b).\par

\begin{figure}[ht]
\centering
\includegraphics[width=\linewidth]{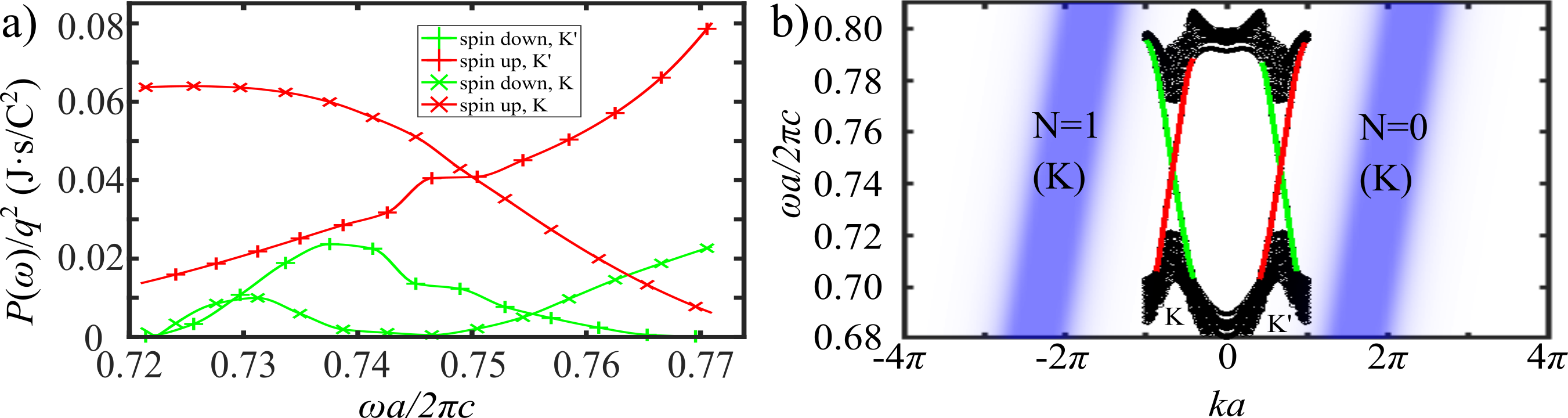}
\caption{(a) The spin- and valley- specific energy spectral density with a charge of $v=0.75c$, where it is clear that edge modes at neither valley is strongly excited. (b)The 1D PBS of the supercell and the strong radiation belts in blue corresponding to Eq. (\ref{eq:kx<0}) with $N=1,0$.}
\label{fig:S3}
\end{figure}

If the quasi-phase matching condition is satisfied for neither valley (i.e. the strong radiation belts cover none of the 4 edge modes), then none of the modes will be strongly excited. For example, if $v=0.75c$ and the beam's path is unchanged (see the strong radiation belts in Fig.~\ref{fig:S3}(b)), then for all 4 modes, $P(\omega)/q^2<0.1$ J$\cdot$s/C$^2$ throughout the bandgap.\par

\begin{figure}[ht]
\centering
\includegraphics[width=\linewidth]{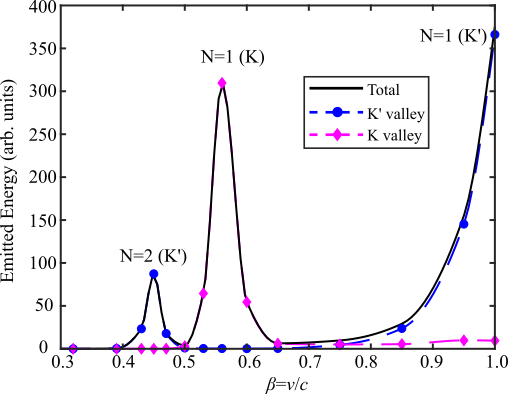}
\caption{TER emitted energy in the topological bandgap $<0.72\omega a/2\pi c<0.77$ as a function of the speed of the charge $v$. Peaks are labelled by the value of $N$ that satisfies in Eqs. (\ref{eq:kx>0}) or (\ref{eq:kx<0}) at that speed.}
\label{fig:E_vs_v}
\end{figure}

The existence of strong QPM indicates that if one plot the total emitted energy in the topological bandgap $<0.72\omega a/2\pi c<0.77$ as a function of the speed of the charge $v$, there would be several peaks corresponding to QPM with different $N$'s. Fig.~\ref{fig:E_vs_v} confirms this, where three quite sharp peaks are evident. The valley-specific energies are also plotted so one can verify that two of the peaks correspond to QPM being satisfied for $K'$ valley modes and one peak corresponds to QPM being satisfied at $K$ valley. This relation between energy and speed of charge is drastically different from the case of excitation of SPPs by electron bombardment of a metal surface where integrated emission energy monotonously increase with electron energy \cite{de2010optical}. The difference emphasizes the role of the lattice plays here, and it is also evident that the mechanism of QPM enhances the TER significantly in this platform.

\section{Experimental setup}
\begin{figure}[ht]
\centering
\includegraphics[width=\linewidth]{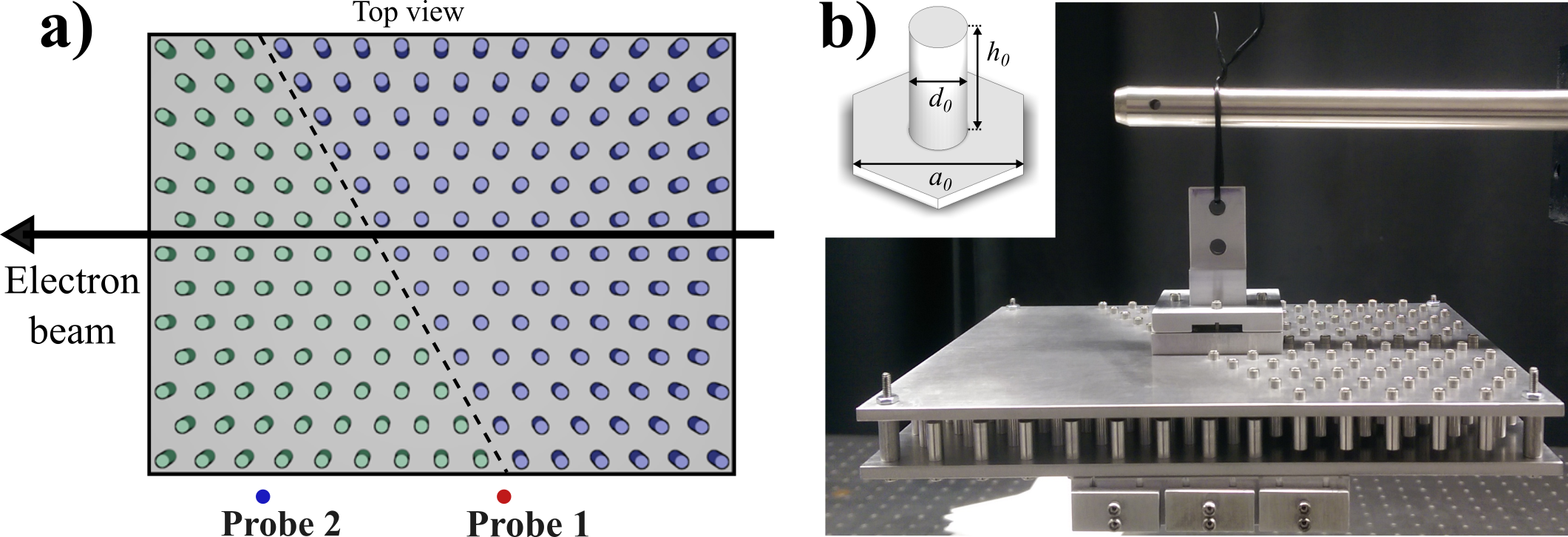}
\caption{(a) Schematic of the 19.5 GHz structure composed of 15$\times$13 QSH-PTIs. Domain with rod attached to the top (bottom) plate is colored in purple (green). Solid dots indicate the position of probes. (b) Fabricated structure with cable holder and counterweight. Inset: geometrical parameters: $a_0$=11.5 mm, $d_0$=3.97 mm, $h_0$=9.78 mm.}
\label{fig:S_exp}
\end{figure}

The operating frequency of the QSH-PTI platform was designed to be around $f_0=19.5$ GHz. (15th harmonics of 1.3 GHz frequency of the photoinjector cavity), with following geometrical parameters: $a_0$=11.5 mm, $d_0$=3.97 mm, $h_0$=9.78 mm as shown in inset of Fig.~\ref{fig:S_exp}. The central operating frequency is chosen for two practical reasons: (i) to maintaining a sizable aperture for an electron bunch to propagate through the structure, and (ii) for the structure to be small enough for the installation into an existing vacuum chamber.

\begin{figure}[ht]
\centering
\includegraphics[width=\linewidth]{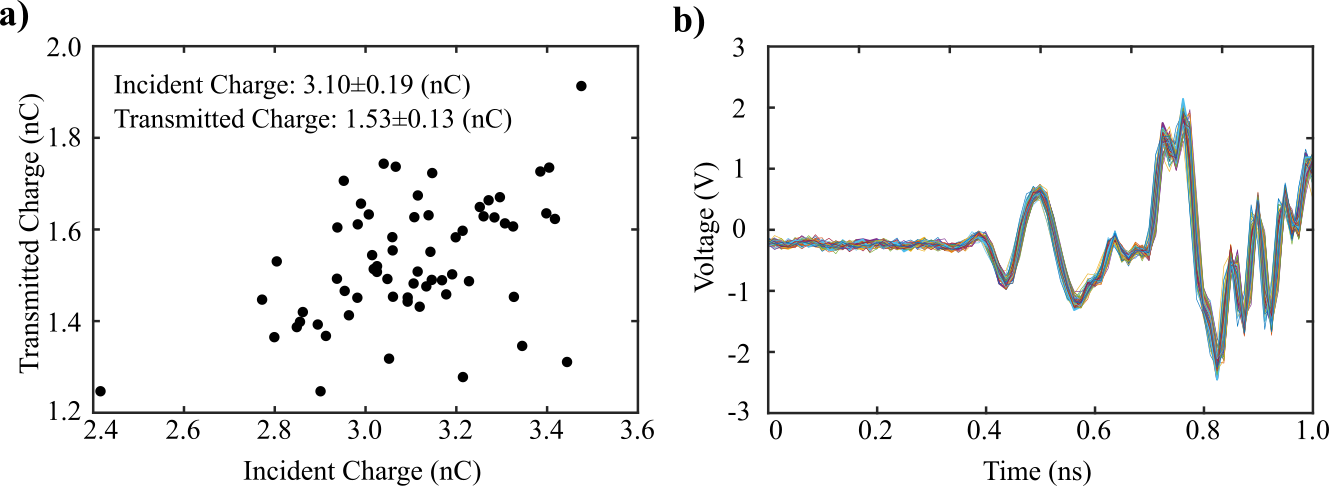}
\caption{Charge and radio frequency measurement for the structure under test. (a) Charge measured by two ICTs on both upstream and downstream side of the experiment chamber. (b) Time domain alignment of acquired radio frequency signal. The figure contains 64 shots of measurements.}
\label{fig:S4}
\end{figure}

A platform consisting of 15$\times$13  QSH-PTI unit cells is considered as the structure under test in this study. This choice of dimensions is a trade-off between limitations such as, available space in vacuum chamber, aperture clearance for electron transmission through structure, and sufficient number of PCs for pronounce collective behaviors. In Fig. \ref{fig:S_exp},
the structure is made of aluminum 6061 with monolithic metal machining and is subsequently assembled with other parts using fasteners. RF probes modified from Kapton insulated soft coaxial cables (30AWG, 50$\Omega$) are clamped by a cable holder and positioned on the longer edge of the structure to measure the edge wave excited by transition radiation of electron as shown in Fig. 5a. A counterweight is attached to the opposite side to render the whole assembly balanced.
The electron beam tests were carried out in collaboration with Argonne Wakefield Accelerator Facility in Argonne National Laboratory (AWA in ANL). The structure under test was installed in a six-way cross vacuum chamber and mounted on an actuator. A 3 nC, 65 MeV  electron bunch with 6 ps  duration was delivered to the vacuum chamber. Charge measurement was taken by two beamline ICTs located both downstream and upstream to the chamber as shown in the Fig. \ref{fig:S4}(a).
The incident charge and transmitted charge were calculated as 3.10$\pm$0.19 nC and 1.53$\pm$0.13 nC respectively which corresponding to 49\% transmission rate. A fast oscilloscope (Keysight DSAZ334A) with 4 channels, 80 GHz sampling rate, and 33 GHz bandwidth was employed to measure the RF signal in this experiment. In Fig. \ref{fig:S4}(b), data from 64 shots were aligned to each other in time domain to ease out shot-to-shot variation for data post-processing.\par
\begin{figure}
\centering
\includegraphics[width=\linewidth]{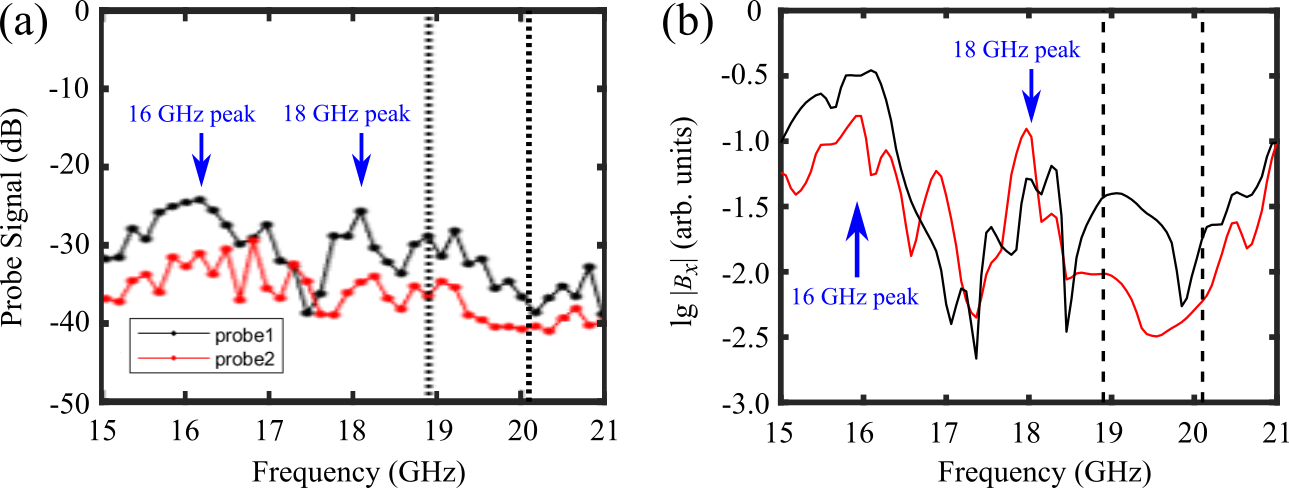}
\caption{Experimentally measured (a) and simulated (b) signals by the two probes, in a wider frequency range 15 to 21 GHz. The two Cherenkov peaks predicted by simulation are clearly visible in the experimental data. Dashed lines mark the boundaries of the photonic bandgap.}
\label{fig:spec}
\end{figure}

In the experiment we also measured two Cherenkov peaks outside but near this topologically non-trivial bandgap, at 16 and 18 GHz, respectively, shown in Fig.~\ref{fig:spec}, as simulation predicts. This confirms our probes were working properly and were sensitive enough to pick up signals at this frequency regime.

\section{TER-based accelerators}
\begin{figure}[ht]
\centering
\includegraphics[width=.9\linewidth]{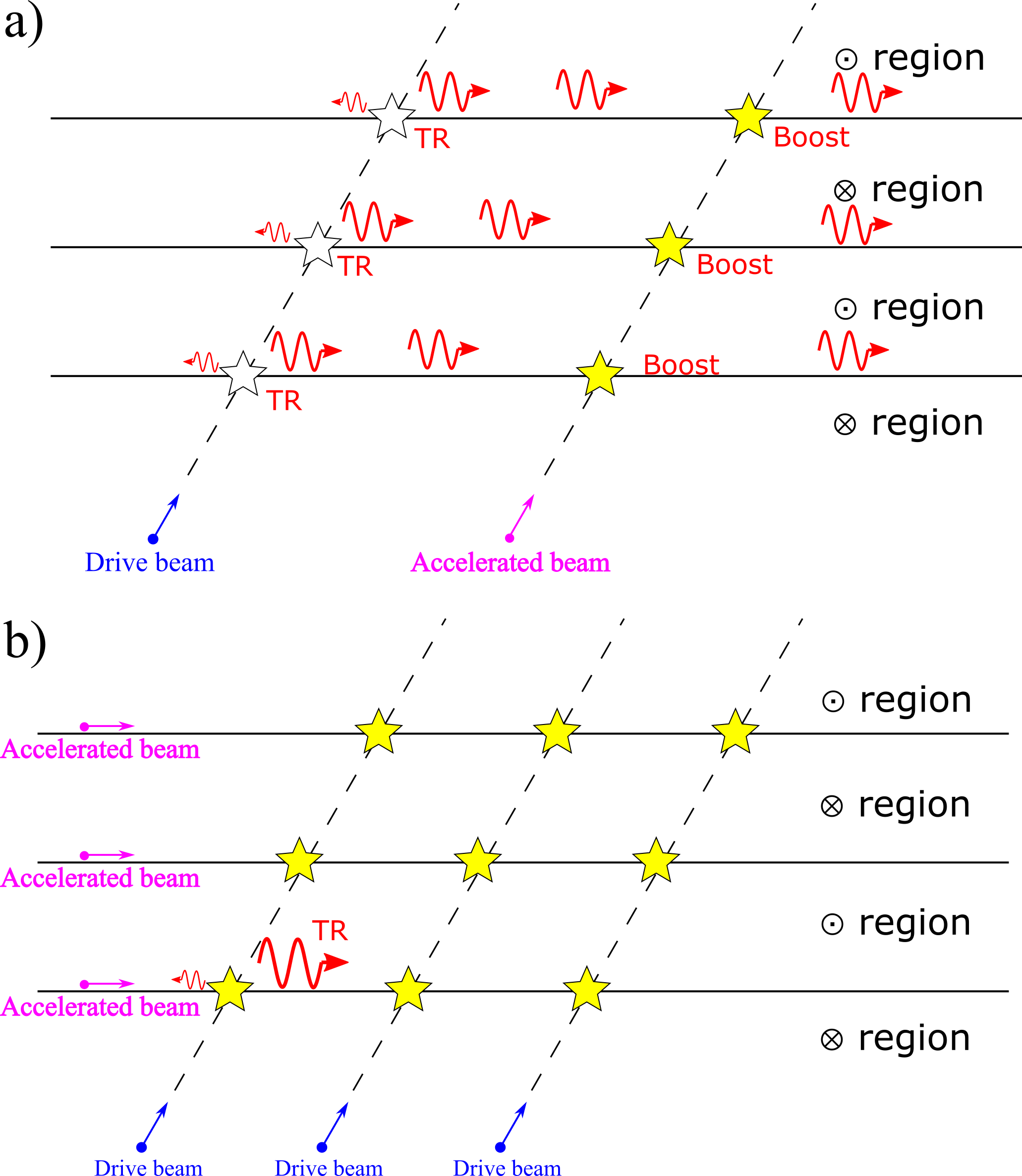}
\caption{\label{fig:acc} (a) A sketch of a TER based two-beam accelerator. It consists of alternative regions of SH-PTIs with rods attached to either the top plate (labeled $\otimes$) or the bottom plate (labeled $\odot$). Individual rods are not plotted. Each time the drive beam crosses the interface (white stars), it will emit TER. The TPEW then propagate along the interface (red arrows), and encounter the accelerated beam, giving it a boost (yellow stars). (b) A sketch of a TER based matrix accelerator.}
\end{figure}

We first show a possible geometry of TER-based non-collinear two-beam accelerators in Fig. \ref{fig:acc}.
It consists of alternative regions of SH-PTIs with rods attached to either the top plate (labeled $\otimes$) or the bottom plate (labeled $\odot$). Each time the drive beam crosses the interface, it will emit TER. The TPEW then propagate along the interfaces, and encounter the accelerated beam, giving it a boost. The accelerated beam does not travel along an interface, so one does not have to worry about synchronization between the beam and TPEWs \cite{whittum1999introduction}, which turns out to be quite difficult. The introduction of TPEWs gives us the privilege to bend the interfaces in the "wave-propagating region", without need to worry about reflection. In practice, a series of bunches with carefully chosen separation can be used to enhance the excitation at a desired frequency. Note according to Fig. 2 in the main text, it's possible to dominantly excite only one spin by wisely choosing the operating frequency. When the driver beam reaches the next interface, by a simple argument of time-reversal symmetry ($\bf\Tilde{J}(r,\omega)\to-\Tilde{J}^*(r,\omega)$, $\mathbf{E}_{k,n}(\mathbf{r})=\mathbf{E}^*_{-k,n}(\bf{r})$, so $|c_n(k,\omega)|\to|c_n(-k,\omega)|$), it will dominantly excite the other spin. However, because of the type of interface also "flips", the radiation still propagate in the same direction. Therefore, it is possible to make the majority of TER produced by the drive beam at every interface all goes towards the accelerated beam, thus increasing the efficiency.\par
Next we show a possible geometry of TER-based matrix accelerators in Fig. \ref{fig:acc}. The differences from the previous design are: (a) now there are multiple drive beams travelling parallel to each other, thus increasing the filed strength; (b) the accelerated beams are not parallel to the drive beams, but instead each travels along an interface; (c) multiple beams can be accelerated at the same time, increasing the overall energy usage efficiency. One challenge for this design is that the interface between two opposite spin-PTIs need to be carefully designed (for example introducing some detours), in order to achieve synchronization between the beam and TPEWs, and high acceleration.
\bibliographystyle{unsrt}
\bibliography{references}